\documentclass["twocolumn"]{webofc}
\usepackage[varg]{txfonts}   % Web of Conferences font
\usepackage{graphics}
\usepackage{graphicx}
\begin{document}
\title{First measurement of the $^{94}$Nb($n$,$\gamma$) cross section at the CERN n\_TOF facility}

\author{
J.~Balibrea-Correa\inst{1} \and %
V.~Babiano-Su\'arez\inst{1} \and %
J.~Lerendegui-Marco\inst{1} \and %
C.~Domingo-Pardo\inst{1} \and %
I.~Ladarescu\inst{1} \and %
A.~Tarife\~no-Saldivia\inst{1} \and %
V.~Alcayne\inst{2} \and %
D.~Cano-Ott\inst{2} \and %
E.~Gonz\'{a}lez-Romero\inst{2} \and %
T.~Mart\'{\i}nez\inst{2}
% ...
%remove \and on previous line
\newline \centering and the n\_TOF Collaboration}
\institute{%
Instituto de F\'{\i}sica Corpuscular, CSIC - Universidad de Valencia, Spain \and
Centro de Investigaciones Energ\'{e}ticas Medioambientales y Tecnol\'{o}gicas (CIEMAT), Spain
% ...
}
\abstract{
One of the crucial ingredients for the improvement of stellar models  is the accurate knowledge of neutron capture cross-sections for the different isotopes involved in the $s$-,$r$- and $i$- processes. These measurements can shed light on existing discrepancies between observed and predicted isotopic abundances and help to constrain the physical conditions where these reactions take place along different stages of stellar evolution.

In the particular case of the radioactive $^{94}$Nb, the $^{94}$Nb($n$,$\gamma$) cross-section could play a role in the determination of the $s$-process production of $^{94}$Mo in AGB stars, which presently cannot be reproduced by state-of-the-art stellar models. There are no previous $^{94}$Nb($n$,$\gamma$) experimental data for the resolved and unresolved resonance regions mainly due to the difficulties in producing high-quality samples and also due to limitations in conventional detection systems commonly used in time-of-flight experiments.

Motivated by this situation, a first measurement of the $^{94}$Nb($n$,$\gamma$) reaction was  carried out at CERN n\_TOF, thereby exploiting the high luminosity of the EAR2 area in combination with a new detection system of small-volume C6D6-detectors and a high quality $^{94}$Nb-sample. The latter was based on hyper-pure $^{93}$Nb material activated at the high-flux reactor of ILL-Grenoble. An innovative ring-configuration detection system in close geometry around the capture sample allowed us to significantly enhance the signal-to-background ratio. This set-up was supplemented with two conventional C$_{6}$D$_{6}$ detectors and  a high-resolution LaCl$_{3}$(Ce)-detector, which will be employed for addressing reliably systematic effects and uncertainties.

At the current status of the data analysis, 18 resonance in $^{94}$Nb+$n$ have been observed for the first time in the neutron energy range from thermal up to 10 keV.
}
\maketitle
\section{Introduction}
\label{intro}
The slow neutron capture process, or $s$ process, is responsible for about half of the isotopic abundances of chemical elements heavier than iron~\cite{Kappeler06}. This process takes place in Massive Stars (>10$M_{\odot}$) and in Asymptotic Giant Branch (AGB) stars of $M$<4$M\textsubscript{\(\odot\)}$ at low or moderated neutron fluxes, mainly produced by the $^{13}C(\alpha,n)$ and $^{22}Ne(\alpha,n)$ reactions\cite{Kappeler11}. In such stellar conditions, the neutron capture time scale is slower than for $\beta$-decay process, thus populating isotopes close to the stability-valley. In order to model and understand the $s$ process and the corresponding stellar media one of the crucial ingredients comes from the nuclear data, in particular from the neutron capture cross-sections of the involved isotopes~\cite{Kappeler11,Herwig05}. This experimental information serves as input for state-of-the-art stellar models~\cite{Kappeler11,Herwig05}, which can be progressively improved and refined by comparing their predictions with astronomical observations of elemental abundances and with isotopic compositions of pre-Solar meteorites~\cite{Zinner98,BURKHARDT11}.

In this context, the isotopic abundances measured in presolar SiC grains represent a long standing puzzle~\cite{BURKHARDT11}. The predicted relative isotopic abundance of $^{94}$Mo is of 1\% or less, whereas isotopic analysis of pre-solar SiC grains yield a factor of five more (5\%)~\cite{Lugaro03}. This problem has been discussed recently by different groups~\cite{Cescutti18,Lugaro03}. One of the possible solutions to this discrepancy could come from the uncertain nuclear data involved in the abundance calculations~\cite{Cescutti18,Lugaro03}. In the particular case of the long-lived $^{94}$Nb isotope, the $^{94}$Nb(n,$\gamma$) cross section could play a role in the $s$-process production of $^{94}$Mo in AGB stars if the $\beta$-decay to $^{94}$Mo would be higher than expected or, correspondingly, the cross section would be smaller~\cite{Cescutti18}. As of today, there are no previous measurements of the $^{94}$Nb($n$,$\gamma$) reaction for the resolved and unresolved resonance regions. This situation was mainly ascribed to difficulties in the production of high-quality $^{94}$Nb samples, as well as limitations in conventional detection systems commonly used in time-of-flight experiments. With the present work, the latter two difficulties have been significantly alleviated, as it is described below.

\section{Experimental challenges and setup}
\label{sec1}

The first challenge in the measurement of the $^{94}$Nb($n$,$\gamma$) cross-section was related to the production of a suitable sample for a TOF experiment. The conventional methodology of producing $^{94}$Nb from activation of commonly available $^{93}$Nb materials is not suitable for an accurate neutron capture cross-section determination based on the time-of-flight technique. Although $^{93}$Nb is naturally mono-isotopic, there is no supplier that can certificate a sample containing less than about 100~ppm of Ta. This quantity is far too much because it would take many years to sit out the $^{182}$Ta produced during a neutron irradiation in a nuclear reactor. To avoid such problematic, a hyper-pure sample of $^{93}$Nb, originally produced at the Institute of Solid State and Materials Research of Dresden~\cite{Koethe00}, was made available for this measurement thanks to a collaboration with the Institute Laue-Langevin (ILL) in France. The synthesis of the ultra-high-purity Nb was based on a refined molten salt electrolysis, zone melting and high and ultra-high vacuum heat treatments. The final $^{93}$Nb material available, 304 mg with $<$1 ppm of Ta, had a shape of two thin wires of 45 and 47 mm long and 1 mm diameter. This high purity material was afterwards activated at the high-flux nuclear ILL-Grenoble reactor for 51 days and a power weighted fluence of 42 full-power days. A careful characterization of the activated sample was performed at the Paul Scherrer Institute (PSI), in Switzerland, by means of a customized HPGe gamma-ray spectroscopy set-up. In this study an activity of 10.1~MBq of $^{94}$Nb was found, without any trace of contaminants, thereby yielding 9.24$\cdot$10$^{18}$ atoms of $^{94}$Nb. This represents $\sim$1\% of the total number of atoms present in the bulk of the sample.

\begin{figure}[h]
\centering
\begin{tabular}{c c}
\includegraphics[width=0.45\columnwidth]{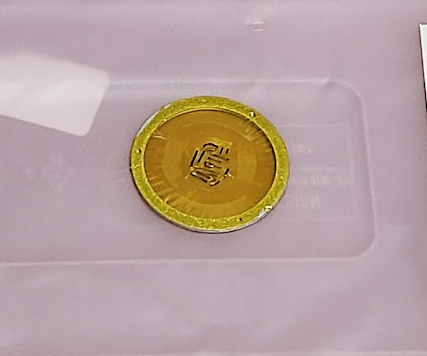} &
\includegraphics[width=0.5\columnwidth]{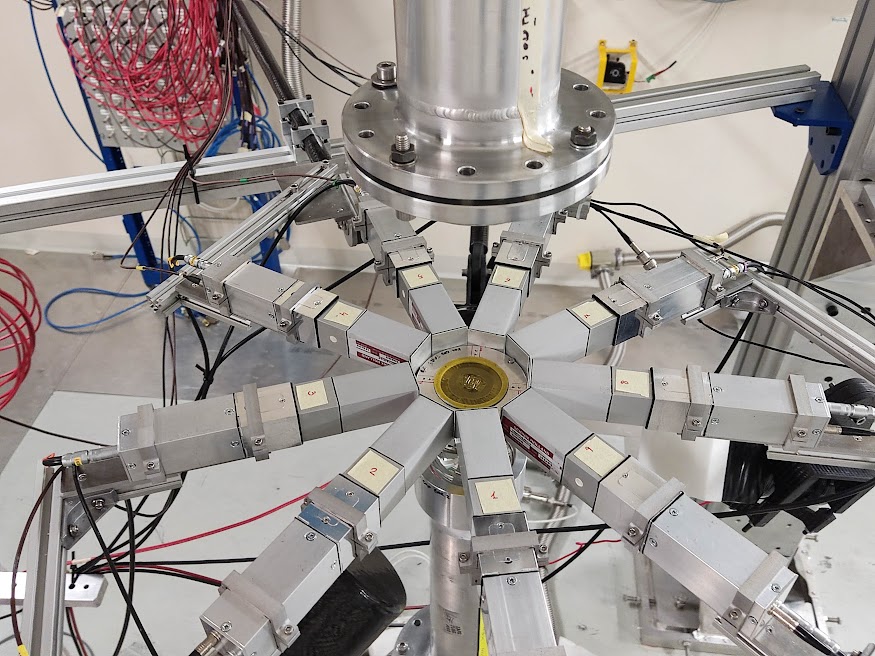}
\end{tabular}
\caption{Left panel: Picture of the $^{94}$Nb sample manufactured at PSI. Right panel: Picture of the experimental setup used during the experimental campaign at the n\_TOF facility in April 2022.}
\label{fig-1}       % Give a unique label
\end{figure}

After the irradiation at ILL and characterization at PSI, the $^{94}$Nb sample for the capture experiment was prepared in a hotcell at PSI. Its final shape is shown in the left panel of Fig~\ref{fig-1}. The initial aim of producing an spiral-shaped sample was hindered by the radiation damage produced in the Nb material during the neutron irradiation at ILL.

The other main challenge in this measurement was related to the decay activity of the Nb-sample itself. The $\beta$-decay of $^{94}$Nb is followed by a two-step $\gamma$-ray cascade of 702~keV and 871 keV. With an activity of 10~MBq, the TOF measurement of this sample is still quite challenging owing to the gamma-ray contribution of the sample activity to the overall background and the relatively small concentration (1\%) of $^{94}$Nb in the sample. Due to the aforementioned arguments, the neutron-capture measurement of $^{94}$Nb was carried out at the vertical experimental area (EAR2) of the CERN n\_TOF facility~\cite{Weiss15}. EAR2 is specially well suited for this particular type of samples because of its high instantaneous neutron flux and its still high energy resolution~\cite{Sabate17}. A full detailed report on the proposal can be found in~\cite{Balibrea21}.
Additionally, during the last years an effort has been made at CERN n\_TOF in order to develop new detection systems capable of coping with the high count-rate requirements of EAR2 and with a fast response to recover quickly after the so-called gamma-flash~\cite{Weiss15,Sabate17}. These requirements have been achieved, to a large extent, with the new segmented Total Energy Detector (s-TED). Their small active volume, which is about 1/9~liter of C$_{6}$D$_{6}$, together with new PMTs (Hamamatsu-R11265U) designed for fast-response and high count-rate conditions, mitigate the effects of the gamma-flash and enlarge the neutron energy range where they can operate under stable and well controlled performance conditions~\cite{Lerendegui23}. A full paper detailing the detector construction, performance and first results is in preparation~\cite{Alcayne23-1}. 
The new s-TED detectors were arranged in an innovative compact-ring configuration, designed to maximize the sensitivity for the ($n$,$\gamma$) reactions in the capture sample itself. A picture of the complete experimental setup during the $^{94}$Nb campaign is shown in the right panel of Fig.~\ref{fig-1}. Thus, the main detection system consists of an array of 9 s-TED detectors arranged as-close-as-possible around the capture sample. Additionally, two C$_{6}$D$_{6}$ and one LaCl$_{3}$(Ce) were placed in the setup with the goal of addressing  systematic effects and uncertainties, detecting possible angular anisotropies in the prompt gamma-ray cascades and extract spectroscopic information. 

The experimental campaign took place between March and April of 2022, including also dedicated runs for background determination and capture-yield normalization. For the latter aim also hyperpure $^{93}$Nb samples and $^{197}$Au samples with specific geometries were used. 

\section{First preliminary results}
\label{sec2}
\begin{figure}[h]
\centering
\begin{tabular}{c c}
\includegraphics[width=0.5\columnwidth]{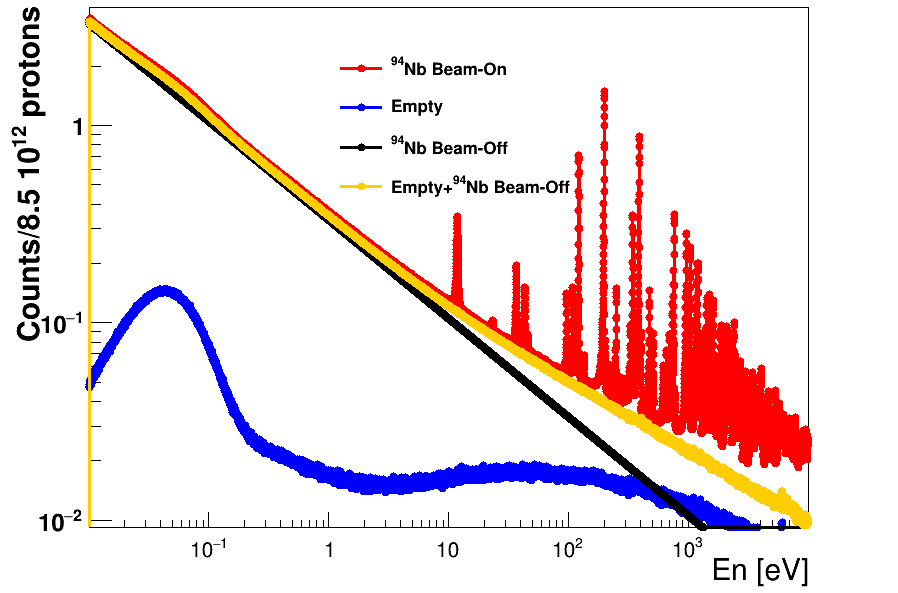} &
\includegraphics[width=0.5\columnwidth]{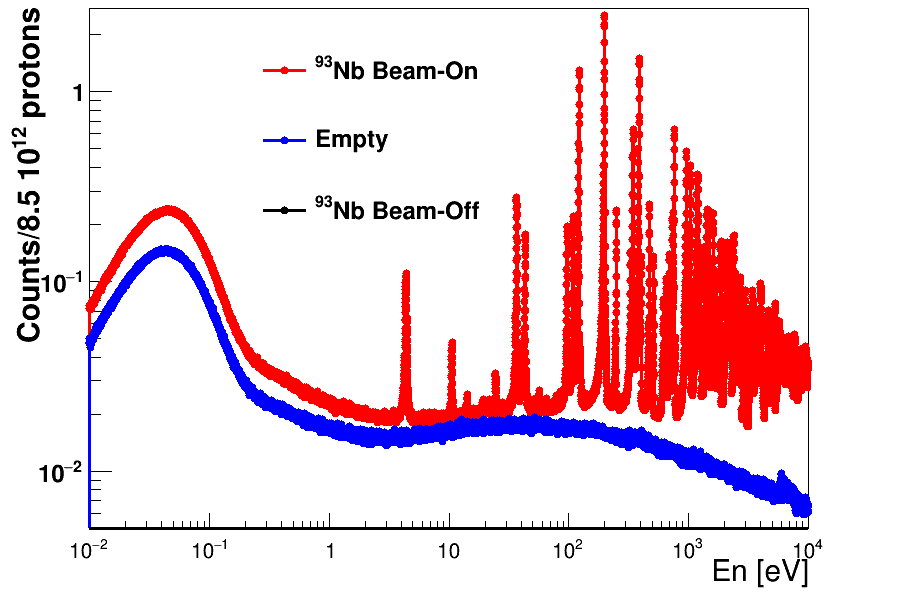}
\end{tabular}
\caption{Counts per proton pulse registered with the s-TED array during different configurations. The left panel shows the results for the $^{94}$Nb-sample measurement, whereas the pure 93Nb sample measurement is shown on the right panel.}
\label{fig-2}       % Give a unique label
\end{figure}

In addition to the main runs with a $^{94}$Nb sample an ancillary measurement of an hyper-pure $^{93}$Nb sample was carried out. The latter had the main aim of determining the contribution of $^{93}$Nb($n$,$\gamma$) within the $^{94}$Nb($n$,$\gamma$) measurement, and also to address corrections for multiple scattering and neutron flux intersection. The counts per proton pulse registered with the s-TED array using the $^{94}$Nb- and $^{93}$Nb-samples in the neutron energy range from thermal up to 10 keV are displayed in both panels of Fig.~\ref{fig-2}. It is worth to mention that, while for the stable $^{93}$Nb-sample the main background component is due to the contribution from the sample backing (blue component in right panel of Fig.~\ref{fig-2}), in the case of the $^{94}$Nb measurement the background is fully dominated by the $\beta$-decays in the radioactive sample itself (black line in the left-panel of Fig.~\ref{fig-2}).

\begin{figure}[h]
\centering
\begin{tabular}{c c}
\includegraphics[width=0.53\columnwidth]{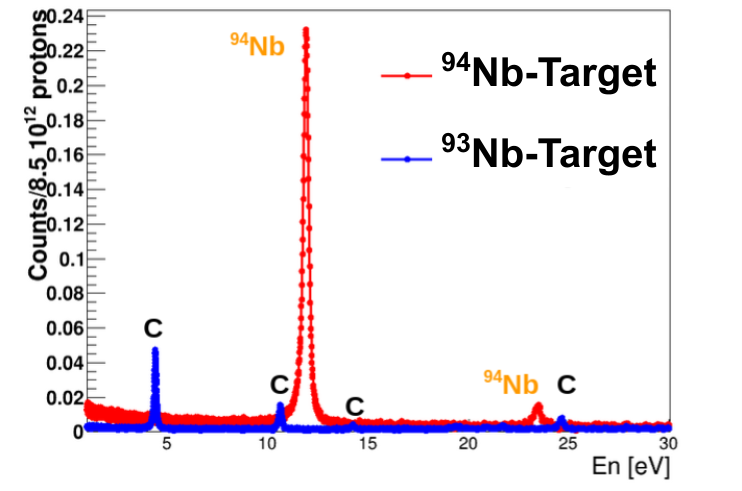} &
\includegraphics[width=0.5\columnwidth]{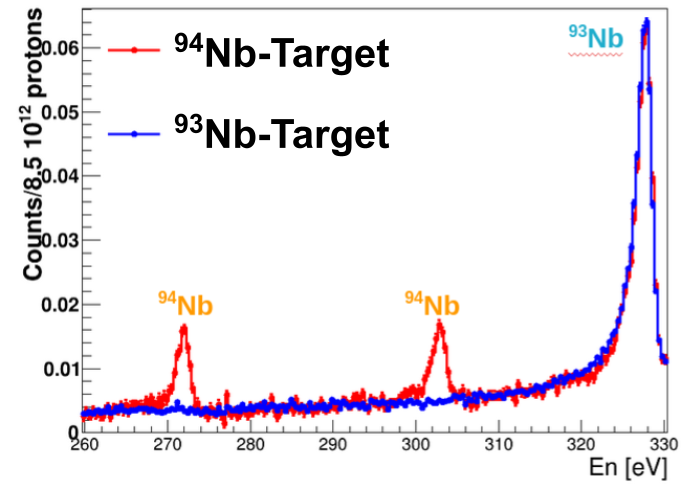}
\end{tabular}
\caption{Comparison between experimental yield after background subtraction for $^{94}$Nb and $^{93}$Nb-spiral configurations for two selected neutron energy ranges. The $^{94}$Nb, $^{93}$Nb and contaminant resonances are tagged by orange, light-blue and black colors, respectively.}
\label{fig-3}       % Give a unique label
\end{figure}

Some preliminary results after background subtraction are shown in both panels of Fig.~\ref{fig-3}.

The left panel in Fig.~\ref{fig-3} shows the first two candidates for $^{94}$Nb+n resonances at low neutron energy. The count-rate of the ancillary $^{93}$Nb-sample measurement is also shown for comparison. Some Ta impurities, labelled as C, are apparent in the latter measurement which, however, do not interfer with the determination of the $^{94}$Nb($n$,$\gamma$) capture yield. 
%In the latest there are some contaminant resonances because of the no hyperpure material used, but they do not represent a problem for cross-section calculation since they do not overlap with any $^{93}$Nb resonance. 
The right-panel in Fig.~\ref{fig-3} shows a higher neutron-energy range, between 260~eV and 330~eV. From the three resonances visible in the figure resonances, only the big resonance near 330~eV can be ascribed to capture on $^{93}$Nb, whereas the other two levels at lower energy belong to capture events on $^{94}$Nb. The observed first neutron resonances up to 200 eV are in very good agreement with the unpublished transmission data from M. R. Serpa performed in 1970~\cite{Serpa:70}.
%In the right panel it is shown another neutron energy range example in which two isolated $^{94}$Nb($n$,$\gamma$) on top of a $^{93}$Nb($n$,$\gamma$) resonance. 
It is worth to remind that the $^{94}$Nb/$^{93}$Nb ratio is of only $\sim$1\%, which shows the overall high sensitivity attained with s-TED detectors for the capture-channel of interest and the large reduction of sample-related background contribution achieved thanks to the high neutron luminosity of EAR2.

At the current preliminary status of the analysis, a completely new capture 18 resonances in the neutron energy range from thermal up to 10 keV have been identified as possible candidates for $^{94}$Nb($n$,$\gamma$) reaction. The next steps in the data analysis comprise the implementation of a Bayesian-based algorithm for reliably addressing the different background components, the determination of the capture-yield by applying the pulse-height weighting technique\cite{Abbondanno04} and the final R-matrix analysis of the resolved resonances.
Additionally it is foreseen to perform a complementary measurement at the n\_TOF NEAR station~\cite{Gervino22} via activation, once the facility is fully characterized and operational, for validating the results from this experimental campaign at several stellar temperatures via activation.

\section{Summary and conclusions}
A first measurement of the $^{94}$Nb($n$,$\gamma$) reaction has been carried out at the CERN n\_TOF EAR2 facility in order to determine the neutron-capture cross-section over a broad neutron-energy range, from thermal energy up to several tens of keV~\cite{Balibrea21}. This experiment was carried out with the aim of resolving discrepancies between AGB-model predictions~\cite{Lugaro03} and measured isotopic abundances of Mo-isotopes in SiC-grains\cite{BURKHARDT11}. The final impact of the measured capture data will be evaluated once the final results of the measurement are obtained.

Until now, there were no previous experimental data on $^{94}$Nb($n$,$\gamma$) owing to the difficulties to produce a suitable sample for a TOF experiment, and the stringent requirements of the experimental set-up for such a TOF measurement.

In this work, a sample containing 304 mg of hyperpure $^{93}$Nb and $^{94}$Nb with an  $^{94}$Nb/$^{93}$Nb isotopic ratio of only $\sim$1\%, was produced thanks to materials and resources from the Institute of Solid State and Materials Research of Dresden, ILL Grenoble  and PSI Villigen.

The $^{94}$Nb($n$,$\gamma$) time-of-flight experiment was carried out at the CERN n\_TOF facility between March and April of 2022 using a new generation of neutron-capture detectors designed to overcome difficulties related to the use of conventional C$_{6}$D$_{6}$ detectors~\cite{Plag03} in the high instantaneous-flux conditions of EAR2. A compact-ring array-configuration around the capture-sample under study allowed one to achieve an excellent signal-to-background ratio and optimize the overall measuring time of 40~days. This set-up was complemented with two conventional C$_{6}$D$_{6}$ detectors and one LaCl$_{3}$(Ce) detector. The latter were aimed at obtaining additional information, that can be relevant to address systematic effects and to derive also spectroscopic information.

The preliminary results reported here show that in this measurement it was possible to cover the neutron energy range from thermal up to several tens of keV. In the present status of the analysis, about 18 capture resonances have been identified as candidates for the $^{94}$Nb($n$,$\gamma$) reaction.

\section*{Acknowledgments}
This work has been carried out in the framework of a project funded by the European Research Council (ERC) under the European Union's Horizon 2020 research and innovation programme (ERC Consolidator Grant project HYMNS, with grant agreement No.~681740). This work was supported by grant ICJ220-045122-I funded by MCIN/AEI/ 10.13039/501100011033 and by European Union NextGenerationEU/PRTR. The authors acknowledge support from the Spanish Ministerio de Ciencia e Innovaci\'on under grants PID2019-104714GB-C21, FPA2017-83946-C2-1-P, FIS2015-71688-ERC, CSIC for funding PIE-201750I26. In line with the principles that apply to scientific publishing and the CERN policy in matters of scientific publications, the
n\_TOF Collaboration recognises the work of V.~Furman and Y.~Kopatch (JINR, Russia), who have contributed to the experiment
used to obtain the results described in this paper.
\bibliography{bibliography}
%
% BibTeX or Biber users please use (the style is already called in the class, ensure that the "woc.bst" style is in your local directory)
% \bibliography{name or your bibliography database}
%\begin{thebibliography}{}
%Journal Author, Journal \textbf{Volume}, page numbers (year)
%Book Author, \textit{Book title} (Publisher, place, year) page numbers
% etc
%\end{thebibliography}

\end{document}